\documentclass[runningheads]{llncs}

\usepackage[T1]{fontenc}
\usepackage{lmodern}
\usepackage{graphicx}
\usepackage{booktabs}
\usepackage{multirow}
\usepackage{amsmath,amssymb}
\usepackage{enumitem}
\usepackage{xspace}
\usepackage{url}
\usepackage{hyperref}
\usepackage{array}
\usepackage{makecell}
\usepackage{comment}

\hypersetup{hidelinks}

\newcommand{\method}{\textsc{TempVideo-MIA}\xspace}

\begin{document}

\title{Membership Inference Attacks Against Video Large Language Models}

\author{
Wei Song\inst{1} \and
Yuxin Cao\inst{2} \and
Ziqi Ding\inst{1} \and
Yi Liu\inst{3} \and
Gelei Deng\inst{3} \and
Yuekang Li\inst{1}
}

\authorrunning{W. Song et al.}

\institute{
UNSW Sydney, Australia\\
\email{\{wei.song1,ziqi.ding1,yuekang.li\}@unsw.edu.au}
\and
National University of Singapore, Singapore\\
\email{yuxincao@u.nus.edu}
\and
Nanyang Technological University, Singapore\\
\email{\{yi009,gelei.deng\}@ntu.edu.sg}
}

\maketitle

\begin{abstract}
Video large language models (VideoLLMs) are increasingly trained or instruction-tuned on large-scale video--text corpora collected from heterogeneous sources, raising an immediate privacy question: can an external auditor determine whether a particular video was used during training? 
While membership inference attacks (MIAs) have been studied extensively for classifiers and, more recently, for text and image generation models, the VideoLLM setting remains unexplored. This setting is challenging because black-box auditors observe only generated text, whereas the membership signal is entangled with video-specific factors such as motion complexity and temporal span. 
In this paper,  we present a black-box MIA targeting VideoLLMs that couples temperature-perturbed generation with video-aware difficulty features. Our key intuition is that member samples tend to induce sharper, more brittle generation behavior across decoding temperatures, and that this signal should be interpreted jointly with the intrinsic difficulty of the queried video. Concretely, we query the target model at low and high temperatures, measure the semantic drift between the resulting texts. We evaluate the attack against \texttt{LLaVA-Video-7B-Qwen2-Video-Only} and achieve a member inference AUC of 0.68 and accuracy of 0.63. These results demonstrate that Video-LLMs are vulnerable to black-box membership inference attacks, highlighting an urgent need for the community to systematically evaluate and mitigate privacy risks in VideoLLMs.

\end{abstract}

\section{Introduction}

Video large language models (VideoLLMs) have rapidly emerged as a powerful interface for video understanding, enabling open-ended captioning \cite{videollama,stylefool,logostylefool}, question answering \cite{llava-video-1,llava-video-2,llava-video-2}, and instruction following over long temporal streams \cite{cao2025failures,VideoSTF,PoisonVID,SecVID,VIDTOKEN}. Recent systems, including LLaVA-Video \cite{llava-video-1,llava-video-2,zhang2024llava-hound,videollama} and Video-ChatGPT \cite{VideoGPT}, combine large language models,  with video encoders and multimodal alignment pipelines to produce natural-language descriptions of complex video content \cite{llava-video-1,llava-video-2,zhang2024llava-hound,videollama,VideoGPT}. 

Despite the rapid advances in VideoLLMs, it also sharpens a rather uncomfortable security question: when a VideoLLM has been instruction-tuned on large and potentially sensitive video--text corpora, can an outside party audit whether a specific sample participated in training? This question matters for at least two reasons. First, many video datasets may contain private, copyrighted, or contractually restricted content. A practical membership inference attack (MIA) offers a mechanism to audit unauthorized data use. Second, MIAs serve as a proxy for memorization and privacy leakage. In language models and other generative systems, memorization has been linked to downstream extraction risks and broader privacy failures \cite{MIA-1,MIA-2,MIA-3}. If analogous leakage exists in VideoLLMs, then the community needs attack methodologies tailored to the multimodal and temporal setting rather than merely inherited from text-only or image-only models.
Unfortunately, direct transplantation is not enough. In a black-box setting, the attacker observes only generated text. The same generation instability can arise either because a sample was memorized or simply because the video itself is intrinsically difficult: highly dynamic motion, long duration, and heavy temporal compression all make the task harder. A useful attack therefore must separate membership-related decoding behavior from video-dependent confounders.

In this paper, we take a first step toward that goal. We introduce \method, a black-box MIA for VideoLLMs built on two simple ideas. First, we probe the model with temperature perturbations. If a queried sample was seen during training, its decoded output may be more tightly concentrated around a memorized mode, making generation more brittle when the decoding temperature changes. We capture this behavior via semantic similarity between low-temperature and high-temperature captions. Second, we explicitly model video difficulty. We use optical-flow-based motion complexity and video duration as difficulty descriptors, and we couple them with the temperature-response signal through interaction features. This yields a compact six-dimensional feature representation that can be consumed by standard binary classifiers.

We evaluate our attack against \texttt{LLaVA-Video-7B-Qwen2-Video-Only}. Member samples are drawn from the publicly reported instruction-tuning corpus \texttt{LLaVA-Video-178K}, while non-members are drawn from the \texttt{VideoChatGPT} dataset. To reduce shortcut learning, we perform length-matched sampling so that member and non-member sets have comparable video-duration distributions. 
Averaged over 100 repeated runs, \method yields an AUC of 0.68 and an accuracy of 0.63 across four classical classifiers: logistic regression (LR) \cite{LR}, random forest (RF) \cite{RF}, support vector machine (SVM) \cite{SVM}, and multilayer perceptron (MLP) \cite{MLP}.

In summary, this paper makes three contributions:
\begin{itemize}[leftmargin=*, topsep=0pt, itemsep=3pt]
    \item We formulate black-box membership inference against VideoLLMs in a practically relevant video captioning setting.
    \item We propose a black-box attack that leverages temperature sensitivity, video-aware difficulty descriptors, and interaction features to infer membership.
    \item We conduct a controlled evaluation under a length-matched protocol and show that VideoLLMs leak measurable membership signals, highlighting a concrete privacy risk in multimodal large models.
\end{itemize}

\section{Background and Related Work}

\subsection{Membership Inferece Attack}
Membership inference attacks ask whether a target sample belonged to a model's training set \cite{MIA-1}. Since the original black-box attacks against classifiers, MIAs have evolved across settings including language models, generative models, and large-scale systems \cite{MIA-1,MIA-2,MIA-3}. In parallel, work on data extraction from language models has shown that memorization can be operationalized into concrete privacy failures \cite{MIA-3}. In practice, MIAs are valuable not only to attackers but also to auditors, dataset owners, and regulators seeking evidence of unauthorized training data use.
Most prior MIA methodology assumes either access to confidence scores, token probabilities, losses, or surrogate shadow models. Those assumptions are often unrealistic for deployed large multimodal systems. Our setting is stricter: the adversary has black-box query access and receives only generated text. This makes the attack weaker in interface but stronger in realism.

Generative models complicate membership inference because outputs are stochastic and there may be no class probabilities to inspect. Prior work has adapted MIAs to generative settings through loss surrogates, reconstruction behavior, or output variance \cite{VID-SME}. More recently, work has started to examine MIAs in large-scale models and label-only settings \cite{VidLeaks,MIA-VISION}. However, the multimodal case---especially the video-language case---remains immature.

The closest intuition behind our method is that stochastic decoding can reveal memorization. If a sample is strongly remembered, perturbing decoding conditions may induce sharper semantic drift than for non-member samples whose outputs are less anchored to a memorized target. We operationalize that intuition without requiring token-level probabilities.

\subsection{Video large language models}
\label{background:videollm}

A typical VideoLLM takes as input a video together with a textual instruction or query, and produces a text response conditioned on both modalities. Its goal is to generate language that reflects the model's semantic understanding of the video under the given prompt. Although implementations differ across systems, most VideoLLMs follow a common architecture consisting of three components: a visual encoder, a modality-alignment module, and a pretrained large language model.
The visual encoder is responsible for transforming raw video frames into compact visual representations. In many recent systems, this encoder is inherited from large-scale vision--language pretraining and instantiated using models such as SigLIP~\cite{zhai2023siglip}, BLIP-2~\cite{li2023blip}, or EVA-CLIP~\cite{fang2023eva}. Since processing every frame in a video is computationally expensive, VideoLLMs usually operate on a sampled subset of frames rather than the full sequence. The extracted visual features are then passed through a projector that maps them into the token embedding space of the language model. Depending on the design, this projector may be implemented as a multi-layer perceptron, a cross-attention module~\cite{vaswani2017attention}, or a query-based transformer such as the Q-Former~\cite{li2023blip}. After this alignment step, the projected visual tokens are concatenated with textual prompt tokens and fed into a pretrained LLM, which serves as the core reasoning and generation module. Common backbones include instruction-tuned language models such as LLaMA~\cite{touvron2023llama}, Vicuna~\cite{chiang2023vicuna}, and Qwen~\cite{bai2023qwen}. This design allows VideoLLMs to process visual and textual information in a unified autoregressive framework, enabling tasks such as video captioning, question answering, and event description.

Formally, let a video be denoted by $\mathcal{V}=\{f_1,f_2,\dots,f_T\}$, where $T$ is the total number of frames and each frame $f_t \in \mathbb{R}^{H \times W \times C}$ has height $H$, width $W$, and $C$ channels. Because full-frame processing is usually infeasible for long videos, the model first samples a subset of $N$ frames from $\mathcal{V}$, where $N \ll T$. We denote the sampled video by
\(
\mathcal{V}_s = \{f_{t_1}, f_{t_2}, \dots, f_{t_N}\}.
\)
Given the sampled frames $\mathcal{V}_s$, the visual encoder $\phi_v$ extracts token-level representations from each frame. Specifically, for each sampled frame $f_{t_i}$, the encoder produces an initial set of $P$ visual tokens. Since directly forwarding all visual tokens to the language model would incur substantial computational cost, these tokens are typically compressed or downsampled to a smaller set of $P'$ tokens, where $P' < P$. Let $\mathbf{v}_i \in \mathbb{R}^{P' \times d_v}$ denote the resulting visual representation for frame $f_{t_i}$, where $d_v$ is the visual embedding dimension. The frame-wise visual features can then be written as
\(
\mathbf{V} = \{\mathbf{v}_1, \mathbf{v}_2, \dots, \mathbf{v}_N\}.
\)
A projector $\phi_p$ subsequently maps these visual features into the token space of the language model:
\begin{equation}
\mathbf{Z} = \phi_p(\mathbf{V}),
\end{equation}
where $\mathbf{Z}$ denotes the projected visual tokens compatible with the LLM input space. Given a text prompt $\mathbf{x}$, the language model $\phi_\ell$ then generates an output response $\mathbf{y}$ conditioned on both the projected visual tokens and the textual input:
\begin{equation}
\mathbf{y} = \phi_\ell(\mathbf{Z}, \mathbf{x}).
\end{equation}

In this way, VideoLLMs implement video understanding by first reducing the raw video into a compact sequence of visual tokens and then performing multimodal reasoning in the language model's embedding space.

Early multimodal large language models (MLLMs), such as Flamingo~\cite{alayrac2022flamingo} and BLIP-2~\cite{li2023blip}, demonstrated that coupling a large language model with a visual encoder can enable strong performance on image-centric tasks, including image captioning and visual question answering~\cite{radford2018improving,touvron2023llama}. Building on this progress, recent research has extended multimodal reasoning from static images to videos, giving rise to VideoLLMs that aim to understand temporally evolving and semantically rich visual content. Unlike image inputs, videos introduce substantial temporal redundancy as well as long-range dependencies across frames, making inference significantly more demanding in both computation and memory. To address this challenge, existing VideoLLMs typically encode sampled video frames into spatio-temporal visual representations, align these representations with the language space, and then pass the resulting embeddings to a pretrained LLM for generation or reasoning~\cite{li2024llava-onevision}.
In practice, however, processing all frames of a video remains prohibitively expensive, especially for long videos. As a result, many representative systems, including Video-LLaMA2~\cite{cheng2024videollama2}, InternVL2.5~\cite{chen2024InternVL2.5}, and NVILA~\cite{liu2025nvila}, rely on sparse uniform frame sampling to keep inference tractable. Prior work such as ViLaMP~\cite{cheng2025vilamp} suggests that sampling 16 frames for shorter videos and 32 frames for longer ones can offer a reasonable efficiency--performance trade-off. Yet this design also introduces an inherent limitation: when the number of sampled frames is fixed while video duration varies, the temporal gap between adjacent sampled frames grows with video length. Consequently, brief but informative events may be missed entirely, and substantial portions of the original video may never be examined. Some methods, including ShareGPT4Video~\cite{chen2024sharegpt4video}, VideoAgent~\cite{fan2024videoagent}, and AKS~\cite{tang2025adaptive}, attempt to alleviate this issue through key-frame selection or adaptive sampling. Even so, the final number of retained frames is still typically small relative to the full video, leaving coverage fundamentally limited.

Beyond frame sampling, VideoLLMs also compress visual tokens to further reduce inference cost. A standard strategy is to downsample token grids after visual encoding, for example by using $2 \times 2$ bilinear interpolation in LLaVA-OneVision~\cite{li2024llava-onevision} and VideoLLaMA3~\cite{videollama}, or average pooling in LLaVA-Video~\cite{zhang2024LLaVA-Video}. Other works adopt more aggressive token reduction mechanisms. For example, LLaMA-VID~\cite{li2024llama-vid} constrains each frame to only two visual tokens, NVILA~\cite{liu2025nvila} increases spatial and temporal resolution before applying pooling-based compression, and Chat-UniVi~\cite{jin2024chat-univi} uses k-nearest-neighbor clustering to merge redundant visual tokens~\cite{jin2024chat-univi}. While these techniques improve efficiency, they inevitably discard part of the original visual signal. In particular, aggressive token compression can remove fine-grained spatial details that may be important for subtle motion understanding, object interactions, or temporally localized events.

\section{Problem Formulation and Threat Model}

\subsection{Problem Formulation}

We study membership inference against a target VideoLLM in the black-box setting. As introduced in Section~\ref{background:videollm}, a VideoLLM takes as input a video $\mathcal{V}$ and a text prompt $\mathbf{x}$, and generates a textual response $\mathbf{y}$. Under decoding temperature $\tau$, the generation process can be written as
\begin{equation}
\mathbf{y}^{(\tau)} = \phi_\ell(\phi_p(\phi_v(\mathcal{V}_s)), \mathbf{x}; \tau),
\end{equation}
where $\mathcal{V}_s \subset \mathcal{V}$ denotes the sampled frame subset processed by the model.
Let $\mathcal{D}_{\mathrm{train}}$ denote the instruction-tuning set used to adapt the target VideoLLM. Each training instance consists of a video and its associated textual supervision, such as a caption or instruction-following response. Given a candidate sample $(\mathcal{V}, \mathbf{t})$, where $\mathcal{V}$ is a video and $\mathbf{t}$ is the associated reference text, the attacker aims to determine whether this sample, or an equivalent supervision instance derived from the same video content, was included in $\mathcal{D}_{\mathrm{train}}$.
We formulate this as a binary inference task over the membership variable:
\begin{equation}
m \in \{0,1\},
\end{equation}
where $m=1$ indicates that the candidate sample is a member of the target model's training data and $m=0$ indicates otherwise. The attacker's objective is to predict $m$ for each candidate sample based on observable signals from the model outputs and auxiliary information extracted from the input video.

\subsection{Threat Model}

We consider a black-box adversary with query access to the target VideoLLM. For a given input video $\mathcal{V}$, the attacker may submit repeated queries using the same prompt $\mathbf{x}$ (e.g., please describe this video) while varying the decoding temperature $\tau$, and observe only the generated textual outputs $\mathbf{y}^{(\tau)}$. The attacker does not have access to token probabilities, logits, hidden states, gradients, losses, model parameters, training metadata, or the target model's training set. In addition to model outputs, the attacker may compute side information directly from the input video itself, such as duration, motion statistics, or other video-level descriptors characterizing input difficulty, as long as these features do not rely on the internals of the target model.

Our threat model assumes that the decoding temperature is exposed as a configurable inference parameter, as is common in many API-based, self-hosted, and evaluation-oriented deployments of generative models. We acknowledge that not all commercial VideoLLM services expose this control to end users; in such cases, our attack characterizes privacy risk under a configurable black-box auditing interface rather than a fully restricted consumer-facing interface. We focus on temperature perturbation because it changes only the stochasticity of decoding while keeping both the input video $\mathcal{V}$ and the semantic query $\mathbf{x}$ fixed. Unlike video-level perturbations, prompt paraphrasing, or frame transformations, varying $\tau$ does not alter the visual content or the intended task, thereby reducing confounding factors introduced by changes in input semantics. We further assume that the attacker has access to an auxiliary shadow dataset drawn from a related video distribution and disjoint from the target model's training set. The shadow dataset is used only to calibrate decision thresholds or train lightweight attack classifiers from black-box observable features.

For each candidate sample, the attacker outputs a binary decision indicating \emph{member} or \emph{non-member}. We evaluate attack performance using standard membership inference metrics, including AUC and classification accuracy. This threat model captures realistic privacy-auditing scenarios for deployed VideoLLMs, where external auditors, dataset owners, enterprise customers, or content providers may interact with a model only through an API, without visibility into its internal states or training corpus, yet still wish to test whether proprietary, private, or sensitive video content was used during training.

\section{Method}
\label{sec:method}

\begin{figure*}[t]
  \centering
  \includegraphics[width=\textwidth]{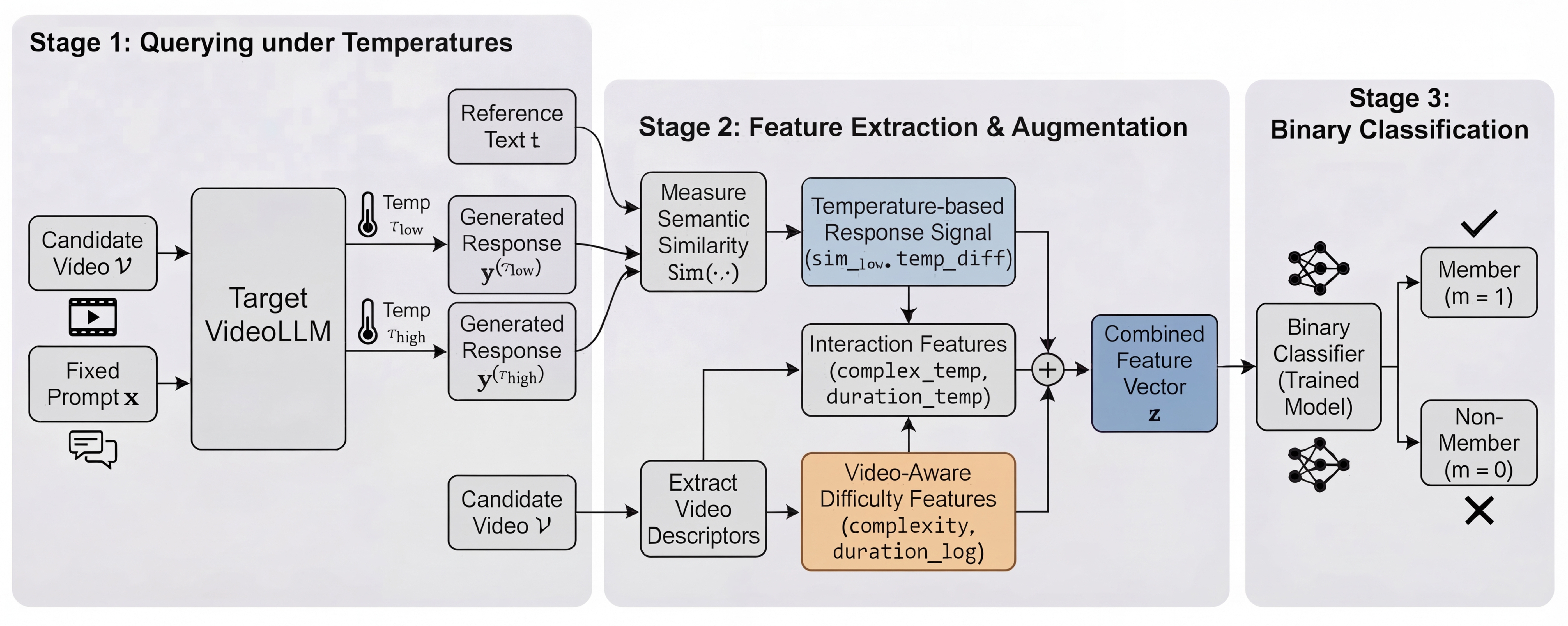}
  \caption{Overview of \method{}.}
  \label{fig:overview}
\end{figure*}

As shown in Figure~\ref{fig:overview}, we introduce a black-box membership inference attack framework, \method, which combines a temperature-based response signal with video-aware difficulty features. At a high level, the attack proceeds in three stages. First, for each candidate video $\mathcal{V}$, we query the target VideoLLM multiple times under different decoding temperatures while keeping the text prompt fixed. Second, we measure how the generated responses change semantically across temperatures and augment this signal with auxiliary descriptors extracted directly from the input video. Third, we train a binary classifier on the resulting feature representation to predict whether the candidate sample is a member of the target model's training set.

Our design is motivated by the hypothesis that membership in VideoLLMs is reflected not only in the content of a single generated response, but also in how the model's response behavior changes under controlled decoding perturbations. In particular, if a video-text pair has been seen during training, the model may form a sharper conditional preference around that sample. As a result, changing the decoding temperature can induce a different response pattern than for unseen samples. At the same time, the video domain introduces an additional complication: response variability is also affected by properties of the input video itself, such as motion and duration. We therefore explicitly model both sources of signal.

\subsection{Temperature-based response signal}

For a candidate video $\mathcal{V}$ and a fixed prompt $\mathbf{x}$, we query the target model under two decoding temperatures, denoted by $\tau_{\mathrm{low}}$ and $\tau_{\mathrm{high}}$. This yields two generated responses,
\begin{equation}
\mathbf{y}^{(\tau_{\mathrm{low}})} = \phi_\ell(\phi_p(\phi_v(\mathcal{V}_s)), \mathbf{x}; \tau_{\mathrm{low}})
\end{equation}
and
\begin{equation}
\mathbf{y}^{(\tau_{\mathrm{high}})} = \phi_\ell(\phi_p(\phi_v(\mathcal{V}_s)), \mathbf{x}; \tau_{\mathrm{high}}).
\end{equation}

To quantify the semantic behavior of these responses, we embed the generated texts into a sentence-level semantic space and compare them against the reference text $\mathbf{t}$. Let $\mathrm{Sim}(\cdot,\cdot)$ denote cosine similarity in the embedding space. We define the low-temperature similarity score as
\begin{equation}
\texttt{sim\_low} = \mathrm{Sim}\!\left(\mathbf{y}^{(\tau_{\mathrm{low}})}, \mathbf{t}\right),
\end{equation}
and analogously define the high-temperature similarity
\begin{equation}
\texttt{sim\_high} = \mathrm{Sim}\!\left(\mathbf{y}^{(\tau_{\mathrm{high}})}, \mathbf{t}\right).
\end{equation}

We then measure temperature sensitivity using the difference
\begin{equation}
\texttt{temp\_diff} = \texttt{sim\_low} - \texttt{sim\_high}.
\end{equation}

The intuition is that member samples may induce a more concentrated conditional generation pattern. Under a lower decoding temperature, generation remains close to that dominant mode, whereas a higher temperature encourages the model to explore alternative phrasings or semantic interpretations, leading to a larger drop in similarity to the reference text. In contrast, for non-member samples, the model's response distribution may already be less concentrated, so the same temperature perturbation can produce a weaker relative change. Under this view, a larger $\texttt{temp\_diff}$ serves as a potential indicator of membership.

\subsection{Video-aware difficulty features}

A temperature-based signal alone is insufficient in the video domain because semantic stability is also shaped by the intrinsic difficulty of the input. Some videos are challenging to caption even when they are familiar to the model, while others are relatively easy even if unseen. To reduce this ambiguity, we augment the attack with auxiliary video descriptors that characterize input difficulty without relying on access to model internals.

\subsubsection{Motion complexity}

We first estimate motion complexity using optical flow statistics. Videos with stronger motion tend to contain richer temporal variation and more rapidly changing visual content, which can make caption generation less stable across repeated queries. Let $\texttt{flow}$ denote the average motion magnitude computed from the input video. We apply a logarithmic transform to reduce scale sensitivity and define
\begin{equation}
\texttt{complexity} = \log(1+\texttt{flow}).
\end{equation}

This feature captures the intuition that highly dynamic videos are often harder for VideoLLMs to summarize consistently, especially when only a subset of frames is sampled by the model.

\subsubsection{Video duration}

We also include video duration as a second difficulty descriptor. Longer videos impose a broader temporal context and increase the chance that important events are missed during sparse frame sampling or subsequent visual token compression. Such inputs may therefore lead to less stable generations, independently of whether the video was seen during training. Let $\texttt{duration}$ denote the video length. We encode it as
\begin{equation}
\texttt{duration\_log} = \log(1+\texttt{duration}).
\end{equation}
Taken together, $\texttt{complexity}$ and $\texttt{duration\_log}$ help distinguish instability caused by genuine input difficulty from instability that is more plausibly associated with non-membership.

\subsection{Interaction features}

Beyond using temperature sensitivity and video descriptors as separate covariates, we further model their interaction explicitly. The reason is that the significance of a temperature-induced semantic shift depends on the difficulty of the underlying video. For example, a large semantic drift may be less surprising for a highly dynamic or very long video than for a short and visually simple one. Conversely, when a difficult video still exhibits a structured and membership-like temperature response, that pattern may be especially informative.
To capture this effect, we construct two interaction terms:
\begin{align}
\texttt{complex\_temp} &= \texttt{complexity} \times \texttt{temp\_diff},\\
\texttt{duration\_temp} &= \texttt{duration\_log} \times \texttt{temp\_diff}.
\end{align}
These features allow the downstream classifier to learn non-linear relationships between memorization-related behavior and video difficulty, rather than treating all temperature shifts as equally informative.

\subsection{Attack representation and classifier}

For each candidate sample, we form a six-dimensional feature vector:
\begin{equation}
\small
\mathbf{z} = 
\texttt{sim\_low}, 
\texttt{temp\_diff}, 
\texttt{complexity}, 
\texttt{duration\_log}, 
\texttt{complex\_temp}, 
\texttt{duration\_temp}.
\end{equation}
We then standardize the features on the training split and fit a binary classifier to predict the membership label $m \in \{0,1\}$. The classifier outputs a membership score for each candidate sample, which is subsequently evaluated using AUC and accuracy.
This design keeps the attack lightweight and fully black-box: it requires only repeated querying of the target VideoLLM, sentence-level embedding comparisons on generated text, and auxiliary statistics computed directly from the input video. At the same time, it is expressive enough to capture both response-level evidence of memorization and video-specific factors that influence output stability in the multimodal setting.

\section{Experimental Setup}

\subsection{Target model and datasets}

We evaluate the attack against \texttt{LLaVA-Video-7B-Qwen2-Video-Only}, a recent VideoLLM associated with the \texttt{LLaVA-Video-178K} instruction-tuning corpus~\cite{llava-video-1}. To construct the membership inference benchmark, we build a balanced evaluation pool containing both member and non-member samples.

The member set consists of 350 video-text samples randomly drawn from \texttt{LLaVA-Video-178K}, which serves as the instruction-tuning dataset associated with the target model. The non-member set consists of 350 samples drawn from the \texttt{VideoChatGPT} dataset~\cite{VideoGPT}, which is disjoint from the target training corpus in our setup. This results in a balanced benchmark of 700 candidate samples in total.
For each sample, we use the video as model input and the paired supervision text as the reference text for feature construction. Unless otherwise specified, all experiments are conducted on this 700-sample benchmark.

\subsection{Attack instantiation}

We instantiate the proposed attack by querying the target model under two decoding temperatures, denoted by $\tau_{\mathrm{low}}$ and $\tau_{\mathrm{high}}$. In our experiments, we set $\tau_{\mathrm{low}}=\text{0}$ and $\tau_{\mathrm{high}}=\text{0.8}$. For each video, we issue the same captioning prompt across temperature settings so that the observed response difference primarily reflects decoding perturbation rather than prompt variation. The prompt used in our experiments is: \texttt{Please describe this video in detail.}

To measure semantic similarity between generated responses and reference text, we embed the texts using CLIP and compute cosine similarity in the resulting embedding space. Following Section~\ref{sec:method}, we extract the temperature-based features $\texttt{sim\_low}$ and $\texttt{temp\_diff}$ from the low- and high-temperature generations.

For the video-aware difficulty descriptors, we compute motion complexity from optical flow and encode video duration using a logarithmic transform. Specifically, optical flow is estimated using \texttt{SpyNet} \cite{SpyNet}, and the average motion magnitude is used to derive $\texttt{complexity}=\log(1+\texttt{flow})$. We then construct the full six-dimensional feature vector described in Section~\ref{sec:method}.

\subsection{Classifiers and evaluation protocol}

We evaluate four standard binary classifiers: logistic regression (LR)~\cite{LR}, random forest (RF)~\cite{RF}, support vector machine (SVM)~\cite{SVM}, and multilayer perceptron (MLP)~\cite{MLP}. For each run, we split the 700-sample benchmark into disjoint training and test sets, standardize features using statistics computed on the training split only, and train the classifier on the resulting attack representation.

To reduce variance from any single data split, we repeat the full pipeline over 100 independent random seeds. Each repetition includes dataset splitting, feature standardization, classifier training, and test-time evaluation. We report the mean AUC, the standard deviation of AUC, and the mean classification accuracy across the 100 runs.

This repeated evaluation protocol is particularly appropriate for a workshop-stage study, as it emphasizes the robustness and stability of the observed attack signal rather than relying on a single favorable split.

\section{Results}

Table~\ref{tab:main-results} summarizes the attack performance across classifiers. All four models achieve AUCs around 0.67, substantially above random guessing. The best mean AUC is obtained by MLP (0.68), followed closely by SVM (0.67) and logistic regression (0.67). The best mean accuracy is also achieved by MLP (0.63).

\begin{table}[t]
\centering
\caption{Membership inference performance over 100 independent runs.}
\label{tab:main-results}
\begin{tabular}{l|c|c|c}
\toprule
Classifier & Mean AUC & Std. AUC & Mean Acc. \\
\midrule
Logistic Regression (LR) & 0.67 & 0.0202 & 0.62 \\
Random Forest (RF)       & 0.67 & 0.0193 & 0.63 \\
Support Vector Machine (SVM) & 0.67 & 0.0204 & 0.63 \\
Multilayer Perceptron (MLP)  & 0.68 & 0.0214 & 0.63 \\
\bottomrule
\end{tabular}
\end{table}

Two observations stand out. First, the signal is consistent across model families. Even linear and margin-based classifiers perform competitively, suggesting that the proposed features encode a stable membership cue. Second, the gain from MLP over the simpler baselines is modest, which is consistent with our design intuition: a small number of carefully chosen features already carries much of the useful signal, while limited non-linearity offers an incremental benefit.

Although the absolute AUC remains moderate, that is not a weakness in itself. In privacy auditing, a reproducible black-box signal above chance is already meaningful, particularly under a constrained interface with no logits, no losses, and no shadow-model assumptions. The result indicates that Video-LLMs may leak training-set membership in ways that are externally testable.

\section{Discussion}

\subsection{Why does the attack work?}

A plausible explanation for the effectiveness of our attack is that instruction-tuned VideoLLMs do not respond to all video-text pairs in the same way. For some training samples, the model appears to develop a sharper conditional preference for particular semantic descriptions, likely due to repeated exposure during instruction tuning. Under low-temperature decoding, generation remains concentrated around this dominant response mode, yielding outputs that stay relatively close to the associated reference text. When the temperature is increased, decoding explores a broader set of candidate continuations, and the resulting semantic drift can become more pronounced. This behavior creates a measurable signal that can be exploited for membership inference.

At the same time, temperature sensitivity alone is not sufficient to explain the observed behavior in the video domain. Unlike text-only or image-only inputs, videos vary substantially in temporal complexity, motion intensity, and duration. These factors directly affect how stably a VideoLLM can summarize or describe a given sample, even in the absence of memorization. For example, a short and visually simple clip may naturally yield consistent generations across repeated queries, whereas a long or highly dynamic video may exhibit larger semantic variation simply because the underlying input is harder to process under sparse frame sampling and visual token compression. This is precisely why video-aware descriptors matter in our setting: they help disentangle response instability caused by input difficulty from response patterns that are more indicative of membership.

The usefulness of the interaction features further supports this interpretation. A large temperature-induced change should not be viewed in isolation; its meaning depends on the type of video on which it occurs. In particular, the same semantic drift may carry different implications for a short, low-motion clip than for a long, high-motion one. By modeling these couplings explicitly, the attack can better capture the fact that memorization-related behavior in VideoLLMs is shaped by both generation dynamics and video-specific difficulty.

\subsection{Security implications}

Our findings suggest that VideoLLMs introduce a practically meaningful privacy risk even in a strictly black-box setting. This is notable because the attacker in our setup does not observe logits, token probabilities, hidden states, gradients, or any other privileged signals commonly used in stronger inference settings. Instead, the attack relies only on generated text responses under controlled decoding perturbations, together with lightweight descriptors computed from the input video. The fact that such limited access is still sufficient to recover non-trivial membership information indicates that privacy leakage in VideoLLMs may be accessible to realistic external parties rather than only to highly privileged insiders.

This has immediate implications for multiple stakeholders. For dataset owners or content providers, our attack provides a concrete auditing tool for testing whether proprietary or sensitive video data may have been used in model training. For model developers and deployers, the results indicate that privacy evaluation for VideoLLMs should not be treated as a simple extension of text-only or image-only membership inference. Temporal structure, frame sampling, and multimodal alignment introduce new sources of both leakage and confounding, which require evaluation protocols tailored to the video-language setting. For the broader research community, our results argue for a more systematic treatment of privacy auditing in multimodal foundation models, including the development of benchmarks, stronger attack methodologies, and principled defenses.

More broadly, the attack surface identified here suggests that standard mitigation ideas such as data deduplication, regularization, output filtering, or differential privacy should not be assumed to transfer directly from other modalities. Their effectiveness must be validated in VideoLLMs specifically, where the interaction between visual compression, temporal sparsity, and language generation may alter both memorization behavior and observable leakage patterns.

\subsection{Why the result matters}

The absolute attack performance in our study is moderate rather than overwhelming, and we view that fact with appropriate caution. An AUC in the high-0.6 range does not imply that membership can be inferred perfectly for every sample, nor does it constitute a complete break of the target model. However, interpreting the result solely through that lens would miss the central point. In a realistic black-box setting, where the attacker observes only generated outputs and has no access to internal confidence signals, consistently outperforming random guessing already provides meaningful evidence that training membership leaves a detectable footprint in model behavior.

This point is strengthened by the design of our evaluation. The observed signal does not arise from a single favorable split or an especially permissive attack setting; rather, it persists across repeated runs and under a feature construction process that explicitly accounts for important video-level confounders such as motion and duration. In that sense, the result is non-trivial not because it yields perfect inference, but because it shows that membership leakage remains measurable even after removing several easy explanations for response variability.

From a security perspective, this is sufficient to raise concern. Privacy failures do not need to be perfect to be consequential. A black-box attack that works reliably above chance can still be operationally useful for auditing, screening, or prioritizing suspicious samples for further investigation. More importantly, the presence of a stable above-chance signal indicates that VideoLLMs are not membership-private by default. That observation alone is enough to motivate stronger privacy evaluations and dedicated defenses for multimodal large models.

\section{Limitations and Future Work}

This study is intentionally preliminary and has several limitations.
First, we evaluate a single target model and a single task setting centered on video captioning. Broader conclusions will require testing across more VideoLLM architectures, scales, and training recipes. Second, while our duration-matching strategy mitigates one obvious shortcut, other dataset-level covariates may remain, such as domain composition, caption style, and content distribution. Third, our current method operates on sentence-embedding similarities rather than token-level likelihoods. This is a deliberate black-box choice, but it may leave signal on the table when richer interfaces are available. Fourth, the current evaluation reports aggregate metrics but does not yet dissect which features contribute most strongly under ablations.
These limitations point directly to promising future directions. A natural next step is a full ablation study that isolates the contribution of temperature perturbation, motion complexity, and interaction features. Another is to examine transferability across target models and prompts. It would also be valuable to test whether repeated multi-sample querying, ensemble scoring, or more structured semantic divergence metrics can push performance higher. On the defense side, future work should explore whether deduplication, instruction-tuning regularization, or privacy-preserving training reduces the attack signal in VideoLLMs.

\section{Conclusion}

We introduced \method, a black-box membership inference attack against VideoLLMs that leverages temperature-induced response variation with video-aware difficulty features. On \texttt{LLaVA-Video-7B-Qwen2-Video-Only}, using a balanced set of member and non-member samples, the attack achieves a best mean AUC of 0.68 across 100 repeated runs. Our results demonstrate that VideoLLMs are not immune to membership inference: even in a restricted black-box setting, training membership leaves a measurable footprint in generated captions.
More broadly, this work highlights that privacy risks in multimodal generative systems cannot be fully understood through methodologies developed for text-only or image-only models. Video introduces distinctive temporal and structural factors that affect both memorization behavior and attack observability. These findings call for stronger privacy auditing benchmarks, more systematic evaluation protocols, and dedicated defenses for VideoLLMs.

\section*{Acknowledgement}
We thank our colleagues and collaborators for valuable discussions and feedback on this work. We are also grateful to the anonymous reviewers for their constructive comments and suggestions, which helped improve the paper. In addition, we thank members of our research group for helpful exchanges on multimodal security, privacy auditing, and VideoLLMs.

\bibliographystyle{splncs04}
\bibliography{ref}

\end{document}